\documentclass[aps,rpl,preprint,groupedaddress]{revtex4}
\begin{document}

\title{Detection of Gravitational Wave - \\
An Application of Relativistic Quantum Information Theory}
\author{Ye Yeo}
\affiliation{Department of Physics, National University of Singapore, 10 Kent Ridge Crescent, Singapore 119260, Singapore}

\author{Chee Leong Ching}
\affiliation{Department of Physics, National University of Singapore, 10 Kent Ridge Crescent, Singapore 119260, Singapore}

\author{Jeremy Chong}
\affiliation{Department of Physics, National University of Singapore, 10 Kent Ridge Crescent, Singapore 119260, Singapore}

\author{Wee Kang Chua}
\affiliation{Department of Physics, National University of Singapore, 10 Kent Ridge Crescent, Singapore 119260, Singapore}

\author{Andreas Dewanto}
\affiliation{Department of Physics, National University of Singapore, 10 Kent Ridge Crescent, Singapore 119260, Singapore}

\author{Zhi Han Lim}
\affiliation{Department of Physics, National University of Singapore, 10 Kent Ridge Crescent, Singapore 119260, Singapore}

\begin{abstract}
We show that a passing gravitational wave may influence the spin entropy and spin negativity of a system of $N$ massive spin-$1/2$ particles, in a way that is characteristic of the radiation.  We establish the specific conditions under which this effect may be nonzero. The change in spin entropy and negativity, however, is extremely small.  Here, we propose and show that this effect may be amplified through entanglement swapping.  Relativistic quantum information theory may have a contribution towards the detection of gravitational wave.
\end{abstract}

\maketitle

Relativity and quantum mechanics are the two pillars of 20th century physics.  Einstein's general relativity \cite{Hartle} is the classical theory of gravity.  Many of its predictions have been experimentally confirmed via very precise measurements.  One of its most intriguing predictions is the propagation of ripples in spacetime curvature at the speed of light ($c = 1$) called {\em gravitational waves}. Gravity is a long-range interaction and it is not possible to shield this interaction.  Gravitational waves thus provide a new window for exploring astronomical phenomena.  However, gravity is the weakest of the four fundamental interactions; this means that gravitational waves are not easily detected.  In fact, they have not yet been detected on Earth.  Quantum mechanics, in combination with computation and information, leads to unexpected new ways that information can be processed and transmitted, extending the known capabilities in the field of classical information to previously unsuspected limits \cite{Nielsen}.  One of the greatest challenges faced by quantum information scientists is the fragility of quantum coherence and entanglement in the presence of environmental decoherence \cite{Zurek}.  In particular, multipartite entangled states such as the GHZ states \cite{Greenberger} become more susceptible under certain kinds of noise as the number of particles increases \cite{Carvalho}.  Motivated by its fundamental importance in gravitational wave detection and several recent developments in relativistic quantum information theory \cite{Peres1}, we explore, in this paper, the possibility of turning this fragility into a quantum means to detect gravitational radiation.

Peres, {\em et al.} \cite{Peres2} were the first to study the relativistic properties of spin entropy for a single, free particle of spin 1/2 and nonzero mass in flat spacetime.  They showed that even if the initial quantum state of the particle is a direct product of a function of momentum and a function of spin, the state under a Lorentz boost is in general not a direct product.  This is because the spin undergoes a Wigner rotation \cite{Wigner} whose direction and magnitude depend on the momentum of the particle.  Spin and momentum appear to be ``entangled''.  As a result, the reduced density matrix for spin becomes mixed and the corresponding entropy becomes nonzero.  Slightly later, Gingrich and Adami \cite{Gingrich} showed that Lorentz boost can also affect the entanglement between spins.  Namely, a maximally entangled Bell state of two massive spin-1/2 particles loses entanglement under a Lorentz boost.

More recently, Terashima and Ueda \cite{Tera1} (see also \cite{Tera2}) extended the original investigation by Peres {\em et al.}, by considering the relativistic quantum mechanics of a massive spin-1/2 particle moving in curved spacetime, which entails a breakdown of the global $SO(3, 1)$ symmetry associated with flat spacetime.  In this case, spin can only be defined locally at each spacetime point by invoking the $SO(3, 1)$ symmetry of some local inertial frame.  Specifically, a spin-1/2 particle in curved spacetime is defined as a particle whose one-particle states furnish the spin-1/2 representation of the {\em local Lorentz transformation}.  Terashima and Ueda showed that, as a consequence of this local definition, the motion of the particle is accompanied by a continuous succession of local Lorentz transformations, which gives rise to spin entropy production that is unique to the curved spacetime.  They illustrated their ideas with the Schwarzschild spacetime in \cite{Tera1, Tera2}.

In this paper, we study the effects on the quantum states of one, two, and in general $N$ spin-1/2 particles due to a plane gravitational wave spacetime propagating in the positive $z$-direction \cite{Hartle}:
\begin{equation}
ds^2 = -dt^2 + [1 + f(t - z)]dx^2 + [1 - f(t - z)]dy^2 + dz^2.
\end{equation}
The size and shape of the propagating ripple in curvature are determined by some dimensionless function $f$ ($|f(t - z)| << 1$).  For example, for a Gaussian wave packet with width $\omega$ and maximum height $A$,
\begin{equation}
f(t - z) = A\exp\left[-\frac{(t - z)^2}{\omega^2}\right].
\end{equation}
And, for a gravitational wave of amplitude $A$ and definite frequency $\varpi$,
\begin{equation}
f(t - z) = A\sin[\varpi(t - z)].
\end{equation}
The only nonvanishing Christoffel symbols for the above metric are $2\Gamma^t_{xx} = -2\Gamma^t_{yy} = \partial f/\partial t$, $-2\Gamma^z_{xx} = 2\Gamma^z_{yy} = \partial f/\partial z$, $\Gamma^x_{tx} = \partial\ln\sqrt{1 + f}/\partial t$, $\Gamma^x_{xz} = \partial\ln\sqrt{1 + f}/\partial z$, $\Gamma^y_{ty} = \partial\ln\sqrt{1 - f}/\partial t$, and $\Gamma^y_{yz} = \partial\ln\sqrt{1 - f}/\partial z$.

We begin with a single spin-1/2 particle $A$ of mass $m$ in a local inertial frame at the spacetime point $x^{\mu}_i$.  This particle is initially prepared at proper time $\tau_i$ in the state
\begin{equation}
|\psi\rangle_A = \int N(k^a)d^3\vec{k}\sum_{\lambda}C(k^a, \lambda)|k^a, \lambda\rangle_A,
\end{equation}
where $N(k^a)d^3\vec{k} = md^3\vec{k}/\sqrt{\vec{k}\cdot\vec{k} + m^2}$ is Lorentz-invariant volume element.  From here on, it is assumed that Latin and Greek letters run over the four inertial-coordinate labels 0, 1, 2, 3 and the four general-coordinate labels, respectively.  $|k^a, \lambda\rangle_A$ is the momentum eigenstate of the particle, labeled by the four-momentum $k^a = (\sqrt{\vec{k}\cdot\vec{k} + m^2}, \vec{k})$ and by the $z$-component $\lambda$ ($= \uparrow$ or 0, $\downarrow$ or 1) of the spin.  We consider, in particular, the case where the coefficient $C(k^a, \lambda) = D(k^a)\delta_{\lambda 0}$,
\begin{equation}
D(k^a) = \frac{1}{\sqrt{N(k^a)}\sqrt{\pi}w}
\prod_{a = 1, 3}\exp\left[-\frac{(k^a - q^a(x_i))^2}{2w^2}\right]\sqrt{\delta(k^2)}.
\end{equation}
We assume that the spacetime curvature does not change drastically within the spacetime scale of the wave packet.  $q^a(x_i)$ is as given in Eq.(18).  Together with the orthogonality condition ${_A}\langle k'^a, \lambda'|k^a, \lambda\rangle_A = \delta^3(\vec{k}' - \vec{k})\delta_{\lambda'\lambda}/N(k^a)$ we clearly have ${_A}\langle\psi|\psi\rangle_A = \int N(k^a)d^3\vec{k}\sum_{\lambda}|C(k^a, \lambda)|^2 = 1$, i.e., $|\psi\rangle_A$ is normalized.  To ease calculations, we set $k^2$ to zero with no loss of generality.  It follows that at $\tau_i$ the reduced density matrix for spin,
\begin{equation}
\rho_A(\tau_i) \equiv \int N(k^a)d^3\vec{k}\ {_A}\langle k^a|\psi\rangle_A\langle\psi|k^a\rangle_A = |0\rangle_A\langle 0|,
\end{equation}
and the corresponding entropy $S_A(\tau_i) \equiv -{\rm tr}[\rho_A(\tau_i)\log_2\rho_A(\tau_i)] = 0$.  We will show that at a later proper time $\tau_f$, $\rho_A(\tau_i)$ evolves to
\begin{equation}
\rho'_A(\tau_f) \equiv {\cal E}[\rho_A(\tau_i)] 
= \frac{1}{2}\left(\begin{array}{cc} 1 + \bar{c} & \bar{s} \\ \bar{s} & 1 - \bar{c}\end{array}\right),
\end{equation}
with spin entropy $S'_A(\tau_f) = -P\log_2P - (1 - P)\log_2(1 - P)$, $P = (1 - |\bar{u}|)/2$.  Here,
\begin{equation}
\bar{u} = \int N(k^a)d^3\vec{k}|D(k^a)|^2\exp(i\Omega),
\end{equation}
with $\Omega = \Omega(k^a; \tau_i, \tau_f, \xi, \vartheta)$ as given in Eq.(22), $\bar{c} = {\rm Re}(\bar{u})$ and $\bar{s} = {\rm Im}(\bar{u})$.  It will be useful to note
\begin{eqnarray}
{\cal E}[R^{00}] \equiv {\cal E}[|0\rangle\langle 0|] 
= \frac{1}{2}\left(\begin{array}{cc} 1 + \bar{c} & \bar{s} \\ \bar{s} & 1 - \bar{c}\end{array}\right), & &
{\cal E}[R^{01}] \equiv {\cal E}[|0\rangle\langle 1|] 
= \frac{1}{2}\left(\begin{array}{cc} -\bar{s} &  1 + \bar{c} \\ -1 + \bar{c} & \bar{s}\end{array}\right), \nonumber \\
{\cal E}[R^{10}] \equiv {\cal E}[|1\rangle\langle 0|] 
= \frac{1}{2}\left(\begin{array}{cc} -\bar{s} & -1 + \bar{c} \\  1 + \bar{c} & \bar{s}\end{array}\right), & &
{\cal E}[R^{11}] \equiv {\cal E}[|1\rangle\langle 1|] 
= \frac{1}{2}\left(\begin{array}{cc} 1 - \bar{c} & -\bar{s} \\ -\bar{s} & 1 + \bar{c}\end{array}\right).
\end{eqnarray}

Next, we consider two spin-1/2 particles $A$ and $B$ with equal mass $m$, initially prepared in the state
\begin{equation}
|\Psi\rangle_{AB} = \int\int N(k^a)N(p^b)d^3\vec{k}d^3\vec{p}\sum_{\lambda, \sigma}C(k^a, \lambda; p^b, \sigma)
|k^a, \lambda\rangle_A \otimes |p^b, \sigma\rangle_B,
\end{equation}
with $C(k^a, \lambda; p^b, \sigma) = D(k^a)D(p^b)\delta_{\lambda\sigma}$.  By writing $|\Psi\rangle_{AB}$ as a density matrix and tracing over the momentum degrees of freedom, we obtain, at $\tau_i$, a maximally entangled Bell state
\begin{equation}
\chi_{AB}(\tau_i) = |\Psi^0_{Bell}\rangle_{AB}\langle\Psi^0_{Bell}| = \frac{1}{2}\sum^1_{j, k = 0} R^{jk}_A \otimes R^{jk}_B,
\end{equation}
where $|\Psi^0_{Bell}\rangle \equiv (|00\rangle + |11\rangle)/\sqrt{2}$, $S_{AB}(\tau_i) \equiv -{\rm tr}[\chi_{AB}(\tau_i)\log_2\chi_{AB}(\tau_i)] = 0$, and (spin) {\em negativity} ${\cal N}[\chi_{AB}(\tau_i)] = 1$.  Consider a density matrix $\chi_{AB}$ and its partial transposition $\chi^{T_A}_{AB}$ for a two spin-1/2 system $AB$.  $\chi_{AB}$ is entangled if and only if $\chi^{T_A}_{AB}$ has any negative eigenvalues \cite{Peres3, Horodecki}.  The negativity \cite{Vidal} is a computable measure of entanglement defined by ${\cal N}[\chi_{AB}] \equiv \max\{-2\sum_i\eta_i,\ 0\}$, where $\eta_i$ is a negative eigenvalue of $\chi^{T_A}_{AB}$.  At $\tau_f$, we will show that
\begin{equation}
\chi'_{AB}(\tau_f) = \frac{1}{2}\sum^1_{j, k = 0} {\cal E}[R^{jk}_A] \otimes {\cal E}[R^{jk}_B]
= \frac{1}{4}\left(\begin{array}{cccc} 
1 + |\bar{u}|^2 & 0                  & 0                  & 1 + |\bar{u}|^2 \\
0               & 1 - |\bar{u}|^2    & -(1 - |\bar{u}|^2) & 0 \\
0               & -(1 - |\bar{u}|^2) & 1 - |\bar{u}|^2    & 0 \\
1 + |\bar{u}|^2 & 0                & 0                    & 1 + |\bar{u}|^2
\end{array}\right),
\end{equation}
with $S_{AB}(\tau_f) = -P\log_2P - (1 - P)\log_2(1 - P)$ but $P = (1 - |\bar{u}|^2)/2$ (see FIG. 1 and 2), and ${\cal N}[\chi'_{AB}(\tau_f)] = |\bar{u}|^2$ (see FIG. 3 and 4).  Generalization to a system of $N$ spin-$1/2$ particles is straightforward:
\begin{eqnarray}
\chi_{A_1\cdots A_N}(\tau_i) & = & |\Psi^0_{GHZ}\rangle_{A_1\cdots A_N}\langle\Psi^0_{GHZ}| 
= \frac{1}{2}\sum^1_{j, k = 0} R^{jk}_{A_1} \otimes \cdots \otimes R^{jk}_{A_N}, \nonumber \\
\chi'_{A_1\cdots A_N}(\tau_f) & = & \frac{1}{2}\sum^1_{j, k = 0} {\cal E}[R^{jk}_{A_1}] \otimes \cdots \otimes {\cal E}[R^{jk}_{A_N}].
\end{eqnarray}
Here, $|\Psi^0_{GHZ}\rangle \equiv (|0\cdots 0\rangle + |1\cdots 1\rangle)/\sqrt{2}$ \cite{Greenberger}.

In order to measure the effects described by Eqs.(7) and (12), we introduce a {\em static} observer at each spacetime point along the ``trajectory'' of the particle(s).  Each observer is assigned a local inertial frame defined by the following convenient choice of {\em vierbein} $e^{\mu}_a(x)$:
\begin{equation}
e^t_0(x) = 1, e^x_1(x) = \frac{1}{\sqrt{1 + f}}, e^y_2(x) = \frac{1}{\sqrt{1 - f}}, e^z_3(x) = 1,
\end{equation}
with all the other components being zero.  Furthermore, we demand that the particle(s) be moving with four-velocity
\begin{equation}
u^{\mu}(x) = (\cosh\xi, \frac{\sinh\xi\sin\vartheta}{\sqrt{1 + f}}, 0, \sinh\xi\cos\vartheta)
\end{equation}
or four-momentum $q^{\mu}(x) = mu^{\mu}(x)$.  Here, $\tanh\xi \equiv v$ ($=$ constant $< 1$), i.e., $\xi$ is the rapidity in the local inertial frame, and $0 < \vartheta < \pi/2$.  In order for the particle(s) to move in this way, which is not a geodesic motion, we must apply an external force.  The acceleration due to this external force is given by $a^{\mu}(x) = u^{\lambda}(x)\nabla_{\lambda}u^{\mu}(x)$:
\begin{equation}
a^{\mu}(x) = 
(\sinh^2\xi\sin^2\vartheta\frac{\partial}{\partial t}\ln\sqrt{1 + f}, \frac{F\sinh\xi\sin\vartheta}{\sqrt{1 + f}}, 0, -\sinh^2\xi\sin^2\vartheta\frac{\partial}{\partial z}\ln\sqrt{1 + f}),
\end{equation}
where
\begin{equation}
F = F(t, z; \xi, \vartheta) \equiv \left(\cosh\xi\frac{\partial}{\partial t} + \sinh\xi\cos\vartheta\frac{\partial}{\partial z}\right)\ln\sqrt{1 + f} = \frac{d}{d\tau}\ln\sqrt{1 + f(t - z)}.
\end{equation}
The inverse of the vierbein $e^a_{\mu}(x)$ in Eq.(14) is given by $e^0_t(x) = 1, e^1_x(x) = \sqrt{1 + f}, e^2_y(x) = \sqrt{1 - f}, e^3_z(x) = 1$.  The vierbein transforms a tensor in a general coordinate system $x^{\mu}$ into that in a local inertial frame $x^a$.  For instance,
\begin{equation}
q^a(x) = e^a_{\mu}(x)q^{\mu}(x) = (m\cosh\xi, m\sinh\xi\sin\vartheta, 0, m\sinh\xi\cos\vartheta),
\end{equation}
and similarly, $a^a(x) = e^a_{\mu}(x)a^{\mu}(x)$ yields
\begin{equation}
a^a(x) = 
(\sinh^2\xi\sin^2\vartheta\frac{\partial}{\partial t}\ln\sqrt{1 + f}, F\sinh\xi\sin\vartheta, 0, -\sinh^2\xi\sin^2\vartheta\frac{\partial}{\partial z}\ln\sqrt{1 + f}).
\end{equation}
A straightforward calculation shows that the nonzero components of the spin connection $\omega^a_{\mu b}(x) \equiv e^a_{\lambda}(x)\nabla_{\mu}e^{\lambda}_b(x)$ are $\omega^0_{x1}(x) = \omega^1_{x0}(x) = \sqrt{1 + f}\partial\ln\sqrt{1 + f}/\partial t$, $\omega^1_{x3}(x) = -\omega^3_{x1}(x) = \sqrt{1 + f}\partial\ln\sqrt{1 + f}/\partial z$, $\omega^0_{y2}(x) = \omega^2_{y0}(x) = \sqrt{1 - f}\partial\ln\sqrt{1 - f}/\partial t$, and $\omega^2_{y3}(x) = -\omega^3_{y2}(x) = \sqrt{1 - f}\partial\ln\sqrt{1 - f}/\partial z$.

Suppose at proper time $\tau$ the particle(s) is at $x^{\mu}$.  After an infinitesimal proper time $d\tau$, the particle(s) moves to a new local inertial frame at the new point $x'^{\mu} = x^{\mu} + u^{\mu}d\tau$.  $q^a(x)$ changes to $q^a(x') = q^a(x) + \delta q^a(x) = \Lambda^a_{\ b}(x)q^b(x)$, where the infinitesimal local Lorentz transformation $\Lambda^a_{\ b}(x) \equiv \delta^a_{\ b} + \lambda^a_{\ b}(x)d\tau$, with $\lambda^a_{\ b}(x) \equiv -[a^a(x)q_b(x) - q^a(x)a_b(x)]/m + \chi^a_{\ b}(x)$ and $\chi^a_{\ b}(x) \equiv -u^{\mu}(x)\omega^a_{\mu b}(x)$.  For our case, we have $\lambda^0_{\ 1}(x) = \lambda^1_{\ 0}(x) =  \sinh^2\xi\cos\vartheta\sin\vartheta G(t, z; \xi, \vartheta)$, $\lambda^0_{\ 3}(x) = \lambda^3_{\ 0}(x) = -\sinh^2\xi\sin^2\vartheta G(t, z; \xi, \vartheta)$, and $\lambda^1_{\ 3}(x) = -\lambda^3_{\ 1}(x) = -\cosh\xi\sinh\xi\sin\vartheta G(t, z; \xi, \vartheta)$; where $G(t, z; \xi, \vartheta) \equiv (\sinh\xi\cos\vartheta\partial/\partial t + \cosh\xi\partial/\partial z)\ln\sqrt{1 + f(t - z)} = -F(t, z; \xi, \vartheta)$.  Corresponding to $\Lambda^a_{\ b}(x)$ is the infinitesimal local Wigner rotation $W^a_{\ b}(x) \equiv \delta^a_{\ b} + \varphi^a_{\ b}(x)d\tau$, where $\varphi^0_{\ 0}(x) = \varphi^0_{\ i}(x) = \varphi^i_{\ 0}(x) = 0$ and $\varphi^i_{\ j}(x) = \lambda^i_{\ j}(x) + [\lambda^i_0(x)k_j - k^i\lambda_{j0}(x)]/(\sqrt{\vec{k}\cdot\vec{k} + m^2} + m)$.  Its spin-1/2 representation is $D^{(1/2)}(W(x)) = \sigma^0 + i[\varphi_{23}(x)\sigma^1 + \varphi_{31}(x)\sigma^2 + \varphi_{12}(x)\sigma^3]d\tau/2$, with the identity matrix $\sigma^0$ and the Pauli matrices $\{\sigma^1, \sigma^2, \sigma^3\}$.  It follows that $\varphi^1_{\ 3}(x) = -G(t, z; \xi, \vartheta)H(k^a; \xi, \vartheta)$, with
\begin{equation}
H(k^a; \xi, \vartheta) \equiv
\left(1 - \frac{k^1\sin\vartheta + k^3\cos\vartheta}{\sqrt{\vec{k}\cdot\vec{k} + m^2} + m}\tanh\xi\right)\cosh\xi\sinh\xi\sin\vartheta.
\end{equation}
Hence, for a finite proper time interval, $\tau_f - \tau_i$, we have
\begin{equation}
D^{(1/2)}(W(x_f, x_i)) = \exp\left[-\frac{i}{2}\sigma^2\Omega(k^a; \tau_i, \tau_f, \xi, \vartheta)\right],
\end{equation}
where
\begin{eqnarray}
\Omega(k^a; \tau_i, \tau_f, \xi, \vartheta) 
& \equiv  & \int^{\tau_f}_{\tau_i}\varphi^1_{\ 3}(x)d\tau \nonumber \\
& =       & H(k^a; \xi, \vartheta)\left[\ln\sqrt{1 + f(t_f - z_f)} - \ln\sqrt{1 + f(t_i - z_i)}\right] \nonumber \\
& \approx & \frac{1}{2}[f(t_f - z_f) - f(t_i - z_i)]H(k^a; \xi, \vartheta).
\end{eqnarray}
Consequently, $|\psi\rangle_A$ evolves to
\begin{equation}
|\psi'\rangle_A = \int N(k^a)d^3\vec{k}\sum_{\lambda, \lambda'}C(k^a, \lambda)
D^{(1/2)}_{\lambda'\lambda}(W(x_f, x_i))|\Lambda(x_f, x_i)k^a, \lambda'\rangle_A,
\end{equation}
and similarly $|\Psi\rangle_{AB}$ to $|\Psi'\rangle_{AB} = \int N(k^a)N(p^b)d^3\vec{k}d^3\vec{p}\sum_{\lambda, \lambda', \sigma, \sigma'}C(k^a, \lambda; p^b, \sigma) \times D^{(1/2)}_{\lambda'\lambda}(W(x_f, x_i))|\Lambda(x_f, x_i)k^a, \lambda'\rangle_A \otimes D^{(1/2)}_{\sigma'\sigma}(W(x_f, x_i))|\Lambda(x_f, x_i)p^b, \sigma'\rangle_B$.  We obtain Eqs.(7) and (12) by writing $|\psi'\rangle_A$ and $|\Psi\rangle_{AB}$ respectively as density matrices, and tracing over the momentum degrees of freedom.  This completes what we set out to do.

In summary, we have shown that the spin entropy of a single massive spin-1/2 particle may change under the influence of a passing gravitational wave.  Interestingly, this change has a dependence on the shape of the wave [see Eqs.(8) and (22)].  In other words, by determining the entropy change, one could in principle deduce $f$.  To measure this change, one could prepare an identical ensemble of many particles in the state $|\psi\rangle$ [Eq.(4)] and subject them to an external force that produces the acceleration in Eq.(16).  The observers at each spacetime point then select a subensemble of particles to determine as accurately as possible its spin state.  The variation of the spin entropy with proper time can then be determined.  We may also consider the same experimental setup for two- or $N$-particle state $|\Psi\rangle$ [Eq.(10) or its generalization, which gives Eq.(13)].  In this case, we can analyze the entanglement properties of the resulting states.  Specifically, we have ${\cal N}[\chi'_{AB}(\tau_f)] = |\bar{u}|^2$.

We have to emphasize that the above effect, even though nonzero, is extremely tiny, especially in the light that the height or amplitude $A$ of a gravitational wave may be of the order of $10^{-21}$.  Consequently, $|\bar{u}|^2$ would be extremely close to $1$.  So, in order to measure such a minute effect, we need to ``amplify'' or ``concentrate'' it.  Our preliminary analysis of the 3- to 7-particle states shows that although a passing gravitational wave may have a greater effect on the 3-particle state compared to a 2-particle one, the 4-, 5-, 6-, and 7-particle states are surprisingly robust.  Thus, it seems, by considering $N$-particle states (with $N \geq 4$) does not help.  Here, we turn to another well-known phenomenon in quantum information science, {\em entanglement swapping} \cite{Zukowski}.  Briefly, we analyze the negativity of the resulting two-particle state
\begin{equation}
\Xi^{(4)}_{A_1A_2} \equiv \frac{1}{p_i}{\rm tr}_{B_1B_2}
[(I_{A_1A_2} \otimes |\Psi^i_{Bell}\rangle_{B_1B_2}\langle\Psi^i_{Bell}|)(\chi'_{A_1B_1}(\tau_f) \otimes \chi'_{A_2B_2}(\tau_f))],
\end{equation}
where $|\Psi^i_{Bell}\rangle = (\sigma^i \otimes \sigma^0)|\Psi^0_{Bell}\rangle$ ($i = 0, 1, 2, 3$) and $p_i = {\rm tr}[(I_{A_1A_2} \otimes |\Psi^i_{Bell}\rangle_{B_1B_2}\langle\Psi^i_{Bell}|) \times (\chi'_{A_1B_1}(\tau_f) \otimes \chi'_{A_2B_2}(\tau_f))] = 1/4$ is the probability of obtaining outcome $i$ from the Bell basis measurement.  Particles $B_1$ and $B_2$ become maximally entangled, but $\Xi^{(4)}_{A_1A_2}$ yields ${\cal N}[\Xi^{(4)}_{A_1A_2}] = |\bar{u}|^4$.  This is therefore an amplification of the decoherence effect due to a gravitational wave.  We repeat the procedure with $\chi'_{A_jB_j}(\tau_f)$ in Eq.(24) replaced by $\Xi^{(4)}_{A_jA_j}$ to obtain $\Xi^{(8)}_{A_1A_2}$, which has negativity ${\cal N}[\Xi^{(8)}_{A_1A_2}] = |\bar{u}|^8$.  It is not difficult to see how one can achieve ${\cal N}[\Xi^{(n)}_{A_1A_2}] = |\bar{u}|^n$, with $n$ the number of particles.  Hence, instead of a direct measurement on the spin states, we subject the particles to the above cycles of entanglement swapping, obtaining a smaller number of pairs of particles with negativities, which differ appreciably from $1$.

In conclusion, we have established the specific conditions under which the spin entropy or negativity of massive spin-$1/2$ particles may change due to a passing gravitational wave.   This very small change may be amplified via the above entanglement swapping scheme, and may be measurable.  It is therefore, a potentially viable means of gravitational wave detection.  More generally, our results demonstrate the exciting possibility of detecting measurable effects due to spacetime curvature using ideas and tools developed in quantum information science.  Effects including those due to our expanding universe will be discussed in a longer paper in preparation \cite{ZhiHan}.

\end{document}